\documentclass[conference,a4paper]{IEEEtran}

\IEEEoverridecommandlockouts

\usepackage[cmex10]{amsmath}
\usepackage{array}
\usepackage{mdwmath}
\usepackage{mdwtab}
\usepackage{eqparbox}
\usepackage{amsmath}
\usepackage{amssymb,amsbsy}
\usepackage{amsthm}
\usepackage{graphicx}
\usepackage{color}

\newcommand{\Fig}[1]{Fig.~\ref{#1}}
\graphicspath{{figures/}}

\newtheorem{theorem}{Theorem}

\newtheorem{remark}{Remark}
\newtheorem{example}{Example}

\newcommand{\F}{\mathbb{F}}

\begin{document}

\sloppy

\title{On a Multiple-Access in a Vector Disjunctive Channel}

\author{
  \IEEEauthorblockN{Alexey Frolov and Victor Zyablov}
	
  \IEEEauthorblockA{\small Inst. for Information Transmission Problems\\
    Russian Academy of Sciences\\Moscow, Russia\\
    Email: \{alexey.frolov, zyablov\}@iitp.ru
  }
  \and
  \IEEEauthorblockN{Vladimir Sidorenko and Robert Fischer\thanks{The work of A.~Frolov has been supported in part by the RFBR grant 12-07-31035.
	V.~Sidorenko is on leave from IITP RAS, Moscow, Russia.}}
  \IEEEauthorblockA{\small Institute of Communications Engineering\\
    University of Ulm\\
    Ulm, Germany \\
    Email: \{vladimir.sidorenko, robert.fischer\}@uni-ulm.de
   }
}




\maketitle
\begin{abstract}
We address the problem of increasing the sum rate in a multiple-access system from \cite{OFZ} for small number of users. We suggest an improved signal-code construction in which in case of a small number of users we give more resources to them. For the resulting multiple-access system a lower bound on the relative sum rate is derived. It is shown to be very close to the maximal value of relative sum rate in \cite{OFZ} even for small number of users. The bound is obtained for the case of decoding by exhaustive search. We also suggest reduced-complexity decoding and compare the maximal number of users in this case and in case of decoding by exhaustive search. 
\end{abstract}

\section{Introduction}
In this paper we consider a noiseless multiuser vector disjunctive (logical OR) channel which is also called Z-channel. This means that ``1'' is always transmitted correctly, and ``0'' may be replaced by ``1''. Let us denote the number of active users by $S$, $S \ge 2$. So for some time $\tau$ the channel inputs are binary vectors ${\bf{x}}_i^{(\tau)}, \: i = 1,2,\ldots, S$, and the channel output at time $\tau$ is an elementwise disjunction of vectors at input
$$
{\bf y}^{(\tau)} = {\bigvee \limits_{i = 1}^{S}}{\bf x}^{(\tau)}_i.
$$

The motivation to consider the channel model is as follows. Consider a communication scheme which uses $M$-ary pulse position modulation (PPM). In this case the channel consists of $M$ subchannels which correspond to either frequencies \cite{A,WLJ} or time slots. To transmit the $i^{\text{th}}$ element of an $M$-ary alphabet the user needs to transmit energy (e.g., a short pulse) in the $i^{\text{th}}$ subchannel. We can say that the $M$-ary symbol is transmitted as a binary vector of length $M$ and weight one. The detector at the receiver measures the energy in the $i^{\text{th}}$ subchannel and decides if ``1'' or ``0'' was transmitted by comparing the energy with the threshold. In this case the probability of receiving ``1'' as ``0'' is much smaller then the probability of receiving ``0'' as ``1'' as we need to suppress the energy in the first case. Thus we use a vector disjunctive channel model as an idealized model for this or similar communication schemes.

The channel model is similar to the A~channel \cite{CW, WZ, BP, VK}, but in the present paper we remove the restriction on the weight of the vectors transmitted by users (in the A~channel model users are only allowed to transmit vectors of weight one).

In \cite{OFZ}, a signal-code construction for the multiple-access system using the A~channel was introduced. The construction was based on Kautz--Singleton (KS) codes \cite{KS}. For the resulting multiple-access system a lower bound on the relative sum rate was derived and it was shown that the bound coincides with an upper bound asymptotically. The signal-code construction requires neither block synchronization nor feedback which are significant advantages of it. All the results were obtained in case of  decoding by exhaustive search. Sure, the decoding algorithm is not applicable in case of codes with large dimensions. In \cite{SF}, a reduced-complexity decoding of KS~codes based on Reed--Solomon (RS) codes was suggested. It was done by a modification of the Guruswami--Sudan list decoder \cite{GS} or the soft-input Koetter--Vardy \cite{KV} decoder of RS~codes. Unfortunately the maximal number of active users was much smaller in comparison to \cite{OFZ}. 

The major disadvantage of the multiple-access system from \cite{OFZ} is as follows. In case of small (in comparison to the number of subchannels) number of users the relative sum rate (or sum rate per subchannel) is very close to zero. Our main goal in the paper is to increase the relative sum rate for a small number of users. To achive this we propose a new signal-code construction in which in case of small number of users we will give more resourses to them. Let the channel consist of $Q$ subchannels. We divide all the range of subchannels into nonoverlapping subranges of $q$ subchannels and give $m = m(S)$ subranges to each user.

Our contribution is as follows. We propose a new signal-code construction, which is an improvement of signal-code construction from \cite{OFZ}. A lower bound on the relative sum rate is obtained and shown to be very close to the maximal value of sum rate in \cite{OFZ} even for small number of users. We find the maximal number of active users in the system. Finally, reduced-complexity decoding is suggested and the maximal number of users is found in this case. 

\section{Basic signal-code construction}

\subsection{Kautz-Singleton Codes}
The Kautz-Singleton code is a concatenation of an outer $(n,k)$-code 
over $\F_q$ and the following inner code. Let us enumerate the elements
of the field $\F_q$ in some order as follows
\[
\F_q = \big\{\, \alpha_1, \alpha_2, \ldots, \alpha_q \,\big\} .
\]
The inner code one-to-one maps every field element $\alpha_i \in \F_q$ to a
binary column vector of length $q$ having a single nonzero element at the
$i^\text{th}$ position. The vector positions are counted from $1$ to $q$.
\begin{example}\label{Ex1}
Let $\textbf{c} = \big[\,\alpha_2,\alpha_4,\alpha_6,\alpha_1,\alpha_2,\alpha_5\,\big]$
be a codeword of $\mathcal{RS}(6,2)$ over $\F_7$. The RS codeword $\textbf{c}$ will
be encoded into a KS codeword $\textbf{C} $ as follows
\[
\textbf{C} = \left[ \begin{array}{cccccc}
    0 & 0 & 0 & 1 & 0 & 0 \\
    1 & 0 & 0 & 0 & 1 & 0 \\
    0 & 0 & 0 & 0 & 0 & 0 \\
    0 & 1 & 0 & 0 & 0 & 0 \\
    0 & 0 & 0 & 0 & 0 & 1 \\
    0 & 0 & 1 & 0 & 0 & 0 \\
    0 & 0 & 0 & 0 & 0 & 0 \\
	\end{array} \right] .
\]
\end{example}

\subsection{Transmission}
Let us recall that the channel consists of $Q$ subchannels. A time interval during which one vector of length $Q$ is transmitted will be called a tact. Assume that all the users use the same alphabet: symbols of $\F_q$. Assume also that each user is given $m$ subranges of $q$ subchannels and the transmission duration (in tacts) is $t$.   

Each user encodes information to be transmitted with the help of a $(n = mt, k, d)$ code $\mathcal C$ (all users use the same code). Consider the process of sending a message by the $i^\text{th}$ user. We denote the codeword to be transmitted by $c_i$. Let $\textbf{C}_i$ be a KS codeword of size $q \times mt$ corresponding to $c_i$. Then $\textbf{C}_i$ is splitted into $m$ disjoint parts $\textbf{C}_i^{(j)},\: j=1,2,\ldots, m$, of size $q \times t$ and each of the parts is sent in the corresponding subrange.

The subranges are allocated dynamically. For this purpose permutations of length $Q$ are used. Permutations used by a current user are known to nobody except for a ``transmitter-receiver'' pair. 

So the transmission can be seen in such a way. The word $\textbf{C}_i$ is splitted into $m$ parts
\[
{{\bf{C}}_i} = \left[ {{{\bf{C}}^{(1)}_i} \: {{\bf{C}}^{(2)}_i} \: \ldots \: {{\bf{C}}^{(m)}_i}} \right]_{q \times mt}.
\]
Then a matrix $\textbf{T}_i$ is formed 
\[
{\textbf{T}_i} = \left[ {\begin{array}{*{20}{c}}
{{{\bf{C}}^{(1)}_i}}\\
{{{\bf{C}}^{(2)}_i}}\\
 \vdots \\
{\begin{array}{*{20}{c}}
{{{\bf{C}}^{(m)}_i}}\\
\textbf{0}
\end{array}}
\end{array}} \right]_{Q \times t}.
\]

Before sending a binary vector, a permutation of its elements is made (a new permutation is used for each vector). In what follows we assume the permutations to be chosen equiprobably and independently from the set of all the $Q!$ possible permutations. 

\subsection{Reception}
The base station sequentially receives messages from all the users. Let us consider the process of receiving a message from the $i^\text{th}$ user. We assume that the base station is synchronized with transmitters of all the users. This means that $t$ columns that correspond to a codeword sent by $i^\text{th}$ user are known at the receiver. At receiving of each column the inverse permutation is performed. Thus, we obtain a matrix
\[
\textbf{Y}_i=\textbf{T}_i \vee \left(\bigvee_{\substack{j=1,\ldots,S\\ j\ne i}}
\textbf{X}_j\right),
\]
where $\textbf{T}_i$ is a matrix corresponding to a KS codeword (${{\bf{C}}_i}$) transmitted by the $i^\text{th}$ user and matrices $\textbf{X}_j, j=1,\ldots,S$, $j\ne i$, are the results of another users activity. Note that matrices $\textbf{X}_j$ may not contain whole codewords sent by another users.

\begin{example}
Let $m = 1$, $Q=q=7$. If, for example, the code matrix $\textbf{C}$ from
\textup{Example \ref{Ex1}} was transmitted then we can receive the following matrix
$\textbf{Y}_i$
\[
\textbf{Y}_i = \left[ \begin{array}{cccccc}
    0 & 0 & 0 & 1 & 0 & 1 \\
    1 & 1 & 1 & 0 & 1 & 1 \\
    0 & 0 & 0 & 0 & 0 & 0 \\
    0 & 1 & 0 & 1 & 0 & 0 \\
    0 & 0 & 0 & 0 & 1 & 1 \\
    0 & 1 & 1 & 0 & 1 & 0 \\
    0 & 0 & 0 & 0 & 0 & 0 \\
	\end{array} \right] .
\]
We see that the channel does not touch the ``$1$''s of the matrix $\textbf{C}$ but replaces 8 zeros by ``$1$''.
\end{example}

The decoding problem is this case is equivalent to decoding problem for KS code. For our basic signal-code construction we use the decoding by exhaustive search. Consider the codeword $c_l\in C$. We need to construct a matrix $\textbf{T}_l$ corresponding to~$c_l$ in the manner described above. Since the described multiple-access system uses a disjunctive channel all the elements of the matrix at the channel output corresponding to the codeword sent by the $i^\text{th}$ user will be non-zero. Therefore, the assumption that the codeword $c_l \in \mathcal C$ was transmitted by the $i^\text{th}$ user is true only if the condition follows
\begin{equation}\label{decode_condition}
\textbf{T}_l \wedge \textbf{Y}_i=\textbf{T}_l,
\end{equation}
where $\wedge$ is an element-wise conjunction of matrices.

To decode we need to check the condition \eqref{decode_condition} for all the words of $\mathcal C$. If the list of codewords satisfying the decoding condition consists of only one word, then the decoder outputs the word; if the list of codewords consists of several words then the decoder outputs a decoding failure (decoding error is not possible in this case).

\begin{remark}
Note, that the presence of a block synchronization is not need here like for the system from \textup{\cite{OFZ}}.
\end{remark}

\begin{remark}
In a real system for permutations to be known both on the transmitter and the receiver it is advisable to use pseudorandom number generators, which are a part of any system based on frequency hopping \textup{\cite{Z}}.
\end{remark}

\subsection{Probability of failure}
Let us estimate the probability $p_*$ of decoding failure for the $i^{\text{th}}$ user. Let $\beta = {1 - {{\left( {1 - \frac{m}{Q}} \right)}^{S - 1}}}$.
\begin{theorem}
\begin{eqnarray*}
p_* \leq \sum\limits_{W = d}^n \left[ {A\left( W \right){\beta^W}} \right] < {q^k}{\beta^d},
\end{eqnarray*} 
where $A(W)$ is the number of codewords of weight $W$ in the code $C$.
\label{p_*}
\end{theorem}
\begin{IEEEproof}
Let the $i^{\text{th}}$ user send a codeword $c_i$, let the matrix $\textbf{T}_i$ correspond to it. The existence of at least one codeword $c^* \ne c_i$ such that the decoding condition follows for a matrix $\textbf{T}^*$ corresponding to it is sufficient for the decoder to output a failure.

Consider some codeword $c' \ne c_i$ and let $D=d(c_i,c')$. The codeword $c'$ will be included in the list of codewords satisfying the decoding condition if and only if a matrix $\textbf{T}'$ is covered by ``$1$''s in $\textbf{Y}_i$.

We can state that al least $n-D$ of $n$ positions to be checked are non-zero as codewords $c'$ and $c_i$ coincide on these positions. Thus a codeword $c'$ will be included in the list if the remaining $D$ positions are non-zero. 

Let us consider one column and let the column contain $l$ of $D$ positions (which should be checked) of a codeword $c'$. Let us enumerate these positions. We are interested in the probability $\textup{P}\left( {\bigcup\nolimits_{j = 1}^l {{A_j}} } \right)$, where $A_j$ is an event consisting in the fact that the position with number $j$ is covered. The probability can be calculated as follows
\[
\textup{P}\left( {\bigcup\limits_{j = 1}^l {{A_j}} } \right) = \textup{P}\left( {{A_1}} \right) \textup{P}\left( {{A_2}|{A_1}} \right) \cdot  \ldots  \cdot \textup{P}\left( {{A_l}|{A_1} \ldots {A_{l - 1}}} \right).
\]

Note that as random independent equiprobable permutations are used
\[
\textup{P}\left( {{A_1}} \right) = \beta,
\]
and
\[
\textup{P}\left( {{A_b|A_1 \ldots A_{b-1}}} \right) \leqslant \beta, \: b=2\ldots l.
\]
Thus, we obtain
\[
\textup{P}\left( {\bigcup\limits_{j = 1}^l {{A_j}} } \right) \leqslant {\beta}^l. 
\]

As a result the probability of a codeword $c'$ to be included in a list is less or equal to $\beta^D$. After getting a sum over all the codewords (except $c_i$) we obtain the needed result.
\end{IEEEproof}

\subsection{Minimal number of tacts}
In this paragraph we estimate the minimal number of tacts $t$ needed to transmit $k$ information symbols with given probability of failure ($p_r$). Let $Q$, $q$, $S$, $m$, $k$ and $p_r$ be fixed.

From Theorem~\ref{p_*} we see that if
\begin{equation}
d \geq \frac{{k{{\log }_2}q - {{\log }_2}{p_r}}}{{ - {{\log }_2}\beta }},
\label{d_ineq}
\end{equation}
then $p_* < p_r$.
We choose the smallest $d$ satisfying (\ref{d_ineq}), i.e.
\begin{equation*}
d = \left\lceil \frac{{k{{\log }_2}q - {{\log }_2}{p_r}}}{{ - {{\log }_2}\beta }} \right\rceil.
\label{d}
\end{equation*}

In accordance to the Gilbert--Varshamov bound if $n$, $d$ and $k$ satisfy the inequality
\[
n \geq k + {\log _q}\left[ {\sum\limits_{i = 0}^{d - 2} { \binom{n-1}{i}{{\left( {q - 1} \right)}^i}} } \right],
\]
then there exist a code with such parameters. Thus, we choose $n$ in such a way 
\begin{equation*}
n = \left\lceil {\frac{{{{\log }_2}q}}{{{{\log }_2}q - 1}}\left( {k + d - 1} \right)} \right\rceil.
\label{n}
\end{equation*}

Now we obtain the estimate on the minimal number of tacts needed in this case:
\begin{eqnarray*}
t &=& \left\lceil {\frac{n}{m}} \right\rceil  \nonumber\\
  &=& \left\lceil {\frac{{{{\log }_2}q}}{{{{\log }_2}q - 1}}\left( {\frac{{k\left( {{{\log }_2}q - {{\log }_2}\beta } \right) - {{\log }_2}{p_r}}}{{m\left( { - {{\log }_2}\beta } \right)}}} \right)} \right\rceil.
\label{t}
\end{eqnarray*}

\subsection{Maximal number of users}
Let $Q$, $q$, $m$, $t$, $k$, $d$ and $p_r$ be fixed. In this paragraph we estimate the maximal number of users for which $p_* < p_r$. Directly from Theorem~\ref{p_*} we obtain
\[
S \le \frac{{ - {{\ln }}\left( {1 - \frac{{\sqrt[d]{{{p_r}}}}}{{{q^{R/\delta }}}}} \right)}}{{ - {{\ln }}\left( {1 - \frac{m}{Q}} \right)}} + 1,
\]
where $\delta = d/n$, $R = k/n$.

And thus
\begin{equation}
S_{\text{max}} \ge \left\lfloor \frac{{ - {{\ln }}\left( {1 - \frac{{\sqrt[d]{{{p_r}}}}}{{{q^{R/\delta }}}}} \right)}}{{ - {{\ln }}\left( {1 - \frac{m}{Q}} \right)}} \right\rfloor + 1.
\label{max_S}
\end{equation}

\begin{remark}
Using the inequalities for logarithm 
\[
\left| x \right| \le  - \ln \left( {1 - \left| x \right|} \right) \le \frac{{\left| x \right|}}{{1 - \left| x \right|}},
\]
we obtain
\[
{S_{\text{max}}} \ge \left\lfloor \left( {\frac{Q}{m} - 1} \right)\frac{{\sqrt[d]{{{p_r}}}}}{{{q^{R/\delta }}}} \right\rfloor + 1.
\]
\end{remark}

\subsection{Sum rate}
Let us introduce some notions. The \textit{rate for one user} (in bits per tact)
\[
{R_i}\left( {Q, q, S, m ,k, p_r} \right) = \frac{k}{t}{\log _2}q.
\]      
                                              
By analogy with the rate of a single user, we define the \textit{sum rate} of all active users as the amount
of information (in bits) transmitted in a system during one tact. Since users transmit information
independently, this value can be computed as follows:
\[
{R_\Sigma } \left( {Q, q, S, m ,k, p_r} \right)  = S\frac{k}{t}{\log _2}q.
\]  

\textit{Relative sum rate} (the rate per subchannel)
\[
{\rho} \left( {Q, q, S, m ,k, p_r} \right)  =  \frac {R_\Sigma}{Q}.
\]  

\begin{figure}[htbp]
\centering
\includegraphics[width=0.4\textwidth]{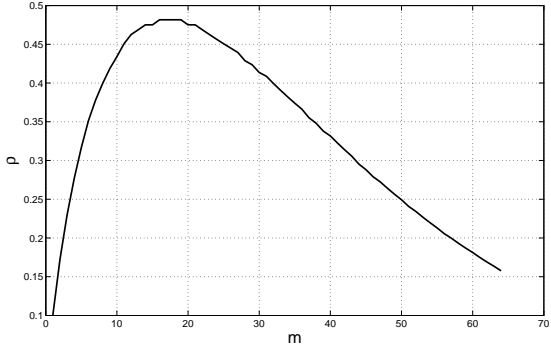}
\caption{Dependency of $\rho$ on $m$}
\label{fig:rho_m}
\end{figure}

In \Fig{fig:rho_m} the dependency of $\rho$ on $m$ is shown. The parameters are chosen as follows: $Q=4096$, $q=64$, $p_r=10^{-10}$, $k = 120$. We see that there is a maximum of relative sum rate at some $m$.

Now define
\[
\rho^* \left( {Q, q, S, k, p_r} \right)  = \mathop {\max }\limits_{1 \le m \le Q/q} \left[ {\rho} \left( {Q, q, S, m ,k, p_r} \right) \right].
\]

\begin{figure}[htbp]
\centering
\includegraphics[width=0.4\textwidth]{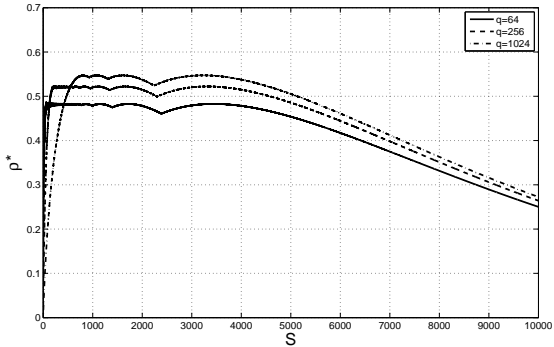}
\caption{Dependency of $\rho^*$ on $S$}
\label{fig:rho_star_nonasymptotic}
\end{figure}

In \Fig{fig:rho_star_nonasymptotic} the dependency of $\rho^*$ on $S$  is shown. The parameters are chosen as follows: $Q=4096$, $p_r=10^{-10}$, $k$ is chosen so that $k \log_2q = 720$ (bits).

Now we derive an asymptotic estimate of a sum rate.

Requiring $p_*$ to decrease exponentially with $n$ (in other words, assuming $p_r = 2^{-cn}$, $c > 0$) and assuming that $k$ is chosen such that $\frac{k}{m} \to \infty$, $\frac{k}{m} = o(Q)$, we obtain
\begin{equation}
t \sim \left( \frac{{{{\log }_2}q}}{{{{\log }_2}q - 1}} \right) \frac{k}{m}\left( {\frac{{{{\log }_2}q - {{\log }_2}\beta }}{{ - {{\log }_2}\beta  - c'}}} \right),
\label{t_equiv}
\end{equation}
where $c' = c \frac{\log_2q}{\log_2(q-1)}$.

\begin{remark}
Note, that the transmission time $t$ here is much better than in \textup{\cite{OFZ}}. Also note, that unlike \textup{\cite{OFZ}} $k$ should be chosen large.
\end{remark}

Let $\mu = m/Q$, $1/Q \leq \mu \leq 1/q$. Let us introduce an asymptotic quantity
\begin{equation}
\rho_\infty(q, S,\mu, k, c)  = \mathop {\lim }\limits_{Q \to \infty } \rho\left( {Q, q, S, m ,k, p_r} \right).
\label{pho_def}
\end{equation}

After substituting (\ref{t_equiv}) into (\ref{pho_def}) we obtain
\begin{eqnarray*}
\rho_\infty \left( {q, S,\mu ,k,c} \right) &\ge & \underline{\rho_\infty} \left( {q,S,\mu, c} \right) \\
&=& S \mu \left( { - {{\log }_2}\beta  - c'} \right)\frac{{{{\log }_2}q - 1}}{{{{\log }_2}q - {{\log }_2}\beta }}.
\end{eqnarray*}

Let us introduce one more quantity
\[
{\underline {\rho^*_\infty}}(q,S,c) = \mathop {\max }\limits_{0 < \mu \leq 1/q} \left\{ \underline {\rho_\infty}  (q,S,\mu, c) \right\}.
\]

\begin{figure}[htbp]
\centering
\includegraphics[width=0.4\textwidth]{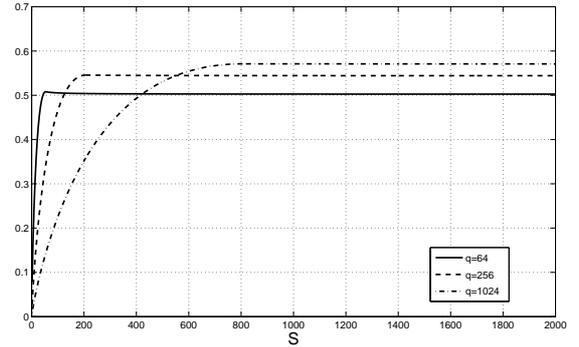}
\caption{Dependency of ${\underline {\rho^*_\infty}}$ on $S$}
\label{fig:rho_star}
\end{figure}

Let $\varepsilon$ be an arbitrarily small positive value, the dependency of $\underline {\rho^*_\infty}(q,S,\varepsilon)$ is shown in \Fig{fig:rho_star}.

Let $\hat \mu  = 1 - \sqrt[{S - 1}]{{\frac{1}{2}}}$, then 
\[
{\underline {\rho^*_\infty}}(q,S,c) \ge \left\{ {\begin{array}{*{20}{l}}
\underline {\rho_\infty} (q,S,\hat \mu ,c), & \quad & \hat \mu  < \frac{1}{q} \\
\underline {\rho_\infty} (q,S,1/q,c),          & \quad &  \text{otherwise}
\end{array}} \right.
\]

\begin{remark}
Note, that the condition $\hat \mu  < \frac{1}{q}$ holds when 
\[
S > \frac{1}{{ - {{\log }_2}\left( {1 - 1/q} \right)}} + 1,
\]
so after strengthening it we obtain 
\[
S > q\ln2 + 1.
\]
\end{remark}

\begin{remark}
Note, that
\[
\underline {\rho_\infty}  (q,S, \hat \mu , \varepsilon)  \ge \left( {1 - \varepsilon '} \right) \frac{{{{\log }_2}q - 1}}{{{{\log }_2}q + 1}} \ln 2,
\]
where $\varepsilon' = S \varepsilon$.
\end{remark}

\section{Reduced-complexity decoding}
As in the proposed scheme $k$ should be chosen large, then the exhaustive search over $q^k$ codewords is not applicable in practice. In this section we propose a reduced-complexity decoding for the signal-code construction.

Let us choose the code $\mathcal C$ of length $n=mt$ to be concatenated, i.e., ${\mathcal C} = {\mathcal C}_O \diamondsuit {\mathcal C}_I$, where ${\mathcal C}_O$ is an outer $(m,k_O,d_O)$-code over $GF(q^{k_I})$, ${\mathcal C}_I$ is an inner $(t,k_I,d_I)$-code over $GF(q)$. 

We will decode the code in such a way. First we will decode all the inner codes independently by exhaustive search (as described above). We choose $k_I$ to be small to have small number of words. Then we will correct the erasures by the outer code. In what follows we assume that the erasure correction is done by means of gaussian elimination (the complexity is $O(m^3)$) to get theoretical results, but in practise it is better to use codes with simple decoding algorithms (e.g., low-density parity-check (LDPC) codes).  

Let us denote by $p^{(I)}_*$ the probability of failure for one inner code. Then in accordance to the Chernoff bound
\[
p_* \le \mathop {\min }\limits_{s > 0} \left\{ {{e^{ - s{d_O}}}\left[ {1 + p_*^{(I)}\left( {{e^s} - 1} \right)} \right]^m} \right\}.
\]

Thus the largest value of $p^{(I)}_*$ for which the requirenment $p_* < p_r$ is satisfied can be calculated as follows
\[
\Hat p_*^{(I)} = \mathop {\max }\limits_{s > 0} \left\{ {\frac{{\sqrt[m]{p_r}{e^{s{\delta_O}}} - 1}}{{{e^s} - 1}}} \right\},
\]
where $\delta_O$ is the relative minimum distance of ${\mathcal C}_O$, i.e., $\delta_O = d_O/m$.

After substituting of $\Hat p_*^{(I)}$ to (\ref{max_S}) we obtain the lower bound on the maximal number of users in case of reduced-complexity decoding
\[
S_{\text{max}} \ge \left\lfloor \frac{{ - {{\ln }}\left( {1 - \frac{{\sqrt[d_I]{{{\Hat p_*^{(I)} }}}}}{{{q^{R_I/\delta_I }}}}} \right)}}{{ - {{\ln }}\left( {1 - \frac{m}{Q}} \right)}} \right\rfloor + 1.
\]

\begin{figure}[htbp]
\centering
\includegraphics[width=0.4\textwidth]{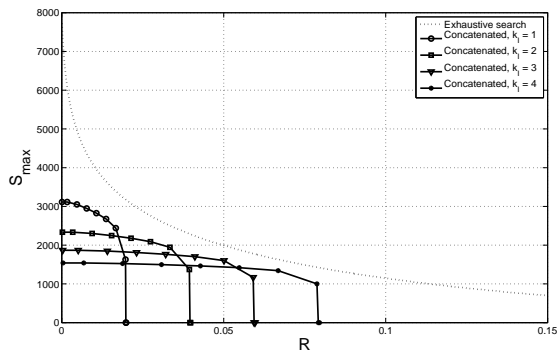}
\caption{Comparison with the decoding by exhaustive search}
\label{fig:conc}
\end{figure}

In \Fig{fig:conc} the dependency of the maximal number of users on the rate of concatenated code is shown for $Q=2^{18}$, $p_r = 10^{-10}$, $m = 200$, $t = 50$, $q = 64$, $k_I = \{1,2,3,4\}$. The comparison with the decoding by exhaustive search is made.

We see that the number of users in case of concatenated construction is smaller in comparison with the decoding by exhaustive search. But at the same time we significantly gain in the complexity of decoding.

\section{Conclusion}
In the present paper, a novel signal-code construction for a multiple-access system using a disjunctive vector channel is proposed. The construction is an improvement of signal-code construction from \cite{OFZ} and it also requires neither block synchronization nor feedback. The main advantage of the signal-code construction in comparison to the construction from \cite{OFZ} is that the sum rate in case of small (in comparison to the number of subchannels) number of users is increased. To achive this we divide all the range of subchannels into nonoverlapping subranges of subchannels and give $m = m(S)$ subranges to each user, where $S$ is the number of users. A lower bound on the relative sum rate for the resulting multiple-access system is obtained and shown to be very close to the maximal value of sum rate in \cite{OFZ} even for small number of users. A lower bound on the maximal number of active users in the system is derrived. These two bounds are obtained for the case of decoding by exhaustive search. Reduced-complexity decoding is suggested. The maximal number of users for reduced-complexity decoding is calculated and compared with the maximal number of users in the case of decoding by exhaustive search.






\end{document}